**Two-Dimensional Moiré Polaronic Electron Crystals**

Eric A. Arsenault[1,*,†], Yiliu Li[1,†], Birui Yang[2], Xi Wang[3,4], Heonjoon Park[3], Edoardo Mosconi[5], Enrico Ronca[6], Takashi Taniguchi[7], Kenji Watanabe[8], Daniel Gamelin[4], Andrew Millis[2], Cory R. Dean[2], Filippo de Angelis[5,6,9,10], Xiaodong Xu[3,11], and X.-Y. Zhu[1,‡]

[1] Department of Chemistry, Columbia University, New York, NY 10027, USA
[2] Department of Physics, Columbia University, New York, NY 10027, USA
[3] Department of Physics, University of Washington, Seattle, WA 98195, USA
[4] Department of Chemistry, University of Washington, Seattle, WA 98195, USA
[5] Istituto CNR di Scienze e Tecnologie Chimiche (CNR-SCITEC), CLHYO, 06123 Perugia, Italy.
[6] Department of Chemistry, Biology and Biotechnology, University of Perugia, 06123, Perugia, Italy
[7] International Center for Materials Nanoarchitectonics, National Institute for Materials Science, Tsukuba, Ibaraki 305-0044, Japan
[8] Research Center for Functional Materials, National Institute for Materials Science, Tsukuba, Ibaraki 305-0044, Japan
[9] Department of Mechanical Engineering, Prince Mohammad Bin Fahd University, Al Khobar, 31952, Saudi Arabia.
[10] SKKU Institute of Energy Science and Technology, Sungkyunkwan University, Suwon, Korea 440-746.
[11] Department of Materials Science and Engineering, University of Washington, Seattle, WA 98195, USA

**ABSTRACT. Two-dimensional moiré materials have emerged as the most versatile platform for realizing quantum phases of electrons. Here, we explore the stability origins of correlated states in $WSe_2/WS_2$ moiré superlattices. We find that ultrafast electronic excitation leads to partial melting of the Mott states on time scales five times longer than predictions from the charge hopping integrals and that the melting rates are thermally activated, with activation energies of 18±3 and 13±2 meV for the one- and two-hole Mott states, respectively, suggesting significant electron-phonon coupling. DFT calculation of the one-hole Mott state confirms polaron formation and yields a hole-polaron binding energy of 16 meV. These findings reveal a close interplay of electron-electron and electron-phonon interactions in stabilizing the polaronic Mott insulators at transition metal dichalcogenide moiré interfaces.**

Realizing quantum phases of electrons with high critical temperatures ($T_c$) has been one of the most important goals in quantum materials research, as exemplified by the longstanding and sometimes contentious quest for high $T_c$ superconductors. Since the discoveries in magic-angle

---

* Junior Fellow in the Simons Society of Fellows, Simons Foundation.
† These authors contributed equally.
‡ To whom correspondence should be addressed. E-mail: xyzhu@columbia.edu.

twisted bilayer graphene (MATBG) [1,2], moiré interfaces have become rich playgrounds for quantum phases. We focus on transition metal dichalcogenide (TMD) moiré superlattices that host a plethora of correlated charge orders [3–12]. The ordered phases are commonly observed at cryogenic temperatures, but a few exhibits high $T_c$, e.g., ~150 K for Mott states [4,5,8,10,12]. While mechanistic discussions have focused on many-body electron-electron (e-e) interactions, the potential role of electron-phonon (e-ph) interactions has been mostly neglected. In the related phenomena of charge density waves (CDWs) [13], e-e and e-ph interactions play equally important roles. Periodic modulation of electron density is accompanied by periodic lattice distortion in a CDW, as one introduces modulation in the potential landscape to the other. Here we explore if periodic electron density in a moiré superlattice also introduces periodic lattice distortion from e-ph coupling

For electrons excited out of an ordered state, the time scale for melting is dictated by that of the electron hopping integral ($t$), $\tau_{e\text{-}e} \sim \hbar/t$, or the period of the relevant phonon mode, $\tau_{ph}$, when the ordering is dominated by e-e or e-ph interactions, respectively [14]. For TMD moiré superlattices hosting correlated states, the hopping integrals are $t$ ~1-5 meV [3,15], corresponding to $\tau_{e\text{-}e}$ ~ 660-130 fs. If e-ph coupling is important, the localization of an electron/hole to a moiré unit cell can be considered as forming a small polaron [16,17]. Deformation of the moiré unit cell likely involves acoustic phonons [16] in momentum range defined by the moiré mini-Brillouin zone, with mean frequencies in the range of ~0.1-1 THz [18] or time scales of ~10-1 ps.

**Correlated Insulator States and Exciton Sensing**

We choose $WSe_2/WS_2$ heterobilayers with twist angles $\theta$ = 0±1° and 60±1°, where correlated electronic states have been demonstrated with exciton sensing [4,5,8,10]. The moiré lattice constants ($a_M$ ~ 8 nm) are the same and band flattening occurs at both twist angles, leading to the similarly observed correlated electronic states. Fig. 1a shows device structure, with an optical image of the 60° device (D1), Fig. 1b, and the reflectance map, Fig. 1c. Under negative (positive) gate bias ($V_g$), holes (electrons) are injected into the heterojunction and, due to type-II band alignment, reside in the $WSe_2$ ($WS_2$) layer [4,5]. We control the moiré lattice filling factor ($\nu$) by $V_g$ and maintain a zero electric field (Methods). Here $\nu < 0$ and $\nu > 0$ correspond to hole and electron doping, respectively. Two additional 0° sample (D2 and D3) are shown in Figs. S1, S2 [19].

The oscillator strength of the sensing exciton is enhanced upon correlated state formation—a result of the accompanying reduction in the effective dielectric constant ($\varepsilon_{eff}$) [4,5,8,10]. This is shown for D1 in the photoluminescence (PL) spectrum from interlayer excitons, Fig. 1d, and reflectance spectrum from intralayer excitons, Fig. 1e. Both spectra as functions of $V_g$ identify the charge neutral ($\nu$ = 0), one- and two-hole ($\nu$ = -1, -2), and one- and two-electron ($\nu$ = 1, 2) states [8,12]. Similar results are obtained for the 0° devices (Fig. S1-S2). While the $\nu$ = -1 state has been called a Mott [20] or a charge-transfer insulator [21], the $\nu$ = -2 state is a Mott insulator of double occupation [22].

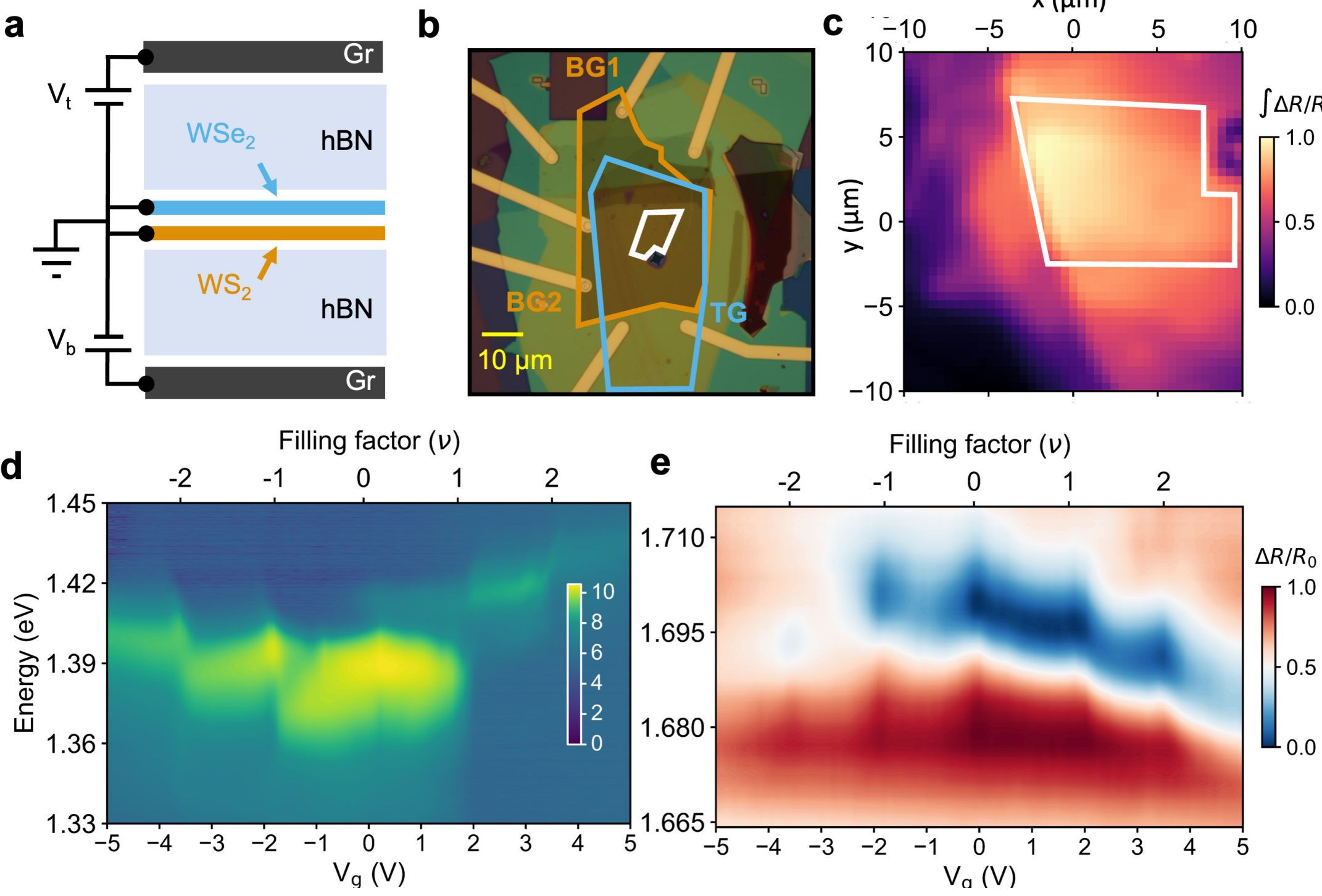

**Figure 1. $WSe_2$/$WSe_2$ device ($\theta$ = 60°, D1): structure and characterization. a** Schematic of the device showing the top ($V_t$) and bottom ($V_b$) gates consisting of few-layer graphite (Gr) and hexagonal boron nitride (hBN) encapsulating the $WSe_2$/$WS_2$ heterobilayer. **b** Optical image of the sample where the top and bottom gates are labeled, along with outlines showing the Gr electrodes (blue and orange lines corresponding to the top and bottom electrodes, respectively) and the $WSe_2$/$WS_2$ overlap region (white line). **c** Steady-state reflectance contrast mapping with the white line highlighting the $WSe_2$/$WS_2$ overlap region. $\Delta R = R - R_0$; $R$ is the reflectance from the sample in the bilayer region and $R_0$ is reflectance from the sample stack without the bilayer. **d** Gated ($V_g = V_t = V_b$) steady-state photoluminescence spectrum (with intensity shown on a logarithmic scale). **e** Gated ($V_g = V_t = V_b$) steady-state reflectance spectrum of the lowest energy moiré exciton of $WSe_2$. All data collected at 11 K.

We apply time-resolved reflectance spectroscopy and focus on the correlated hole states, because sensing from the $WSe_2$ moiré exciton is most sensitive to holes residing in the same layer. A correlated hole Mott state can be represented by gap opening near the top of the valence band, resulting in a lower Hubbard band (LHB, with holes) and an upper Hubbard band (UHB, without holes). We choose a pump photon energy (1.55 eV) below the optical gaps of $WSe_2$ and $WS_2$ to avoid long-lived interlayer excitons [23]. At this photon energy, a hole in the LHB can be excited deep into the valence band; the excited hole cools on ultrafast time scales to reach the UHB, Fig. 2a. The e-h pair across the Hubbard gap, Fig. 2b, is usually called a holon-doublon pair [13,14]. The subsequent disordering may lead to partial gap closing, Fig. 2c, with accompanying increase in $\varepsilon_{eff}$. Figs. 2d-f show 2D plots of transient reflectance as a function $V_g$ at selected delays from the 60° device; the signal is $\Delta R/R$, where $\Delta R$ is the pump-induced reflectance change and $R$ is the reflectance without pump. The peaks in $\Delta R/R$ only appear around $V_g$ values commensurate with $\nu$ = -1 and -2 and vary with Δt. Similar results are obtained for the 0° samples (Fig. S1-S2).

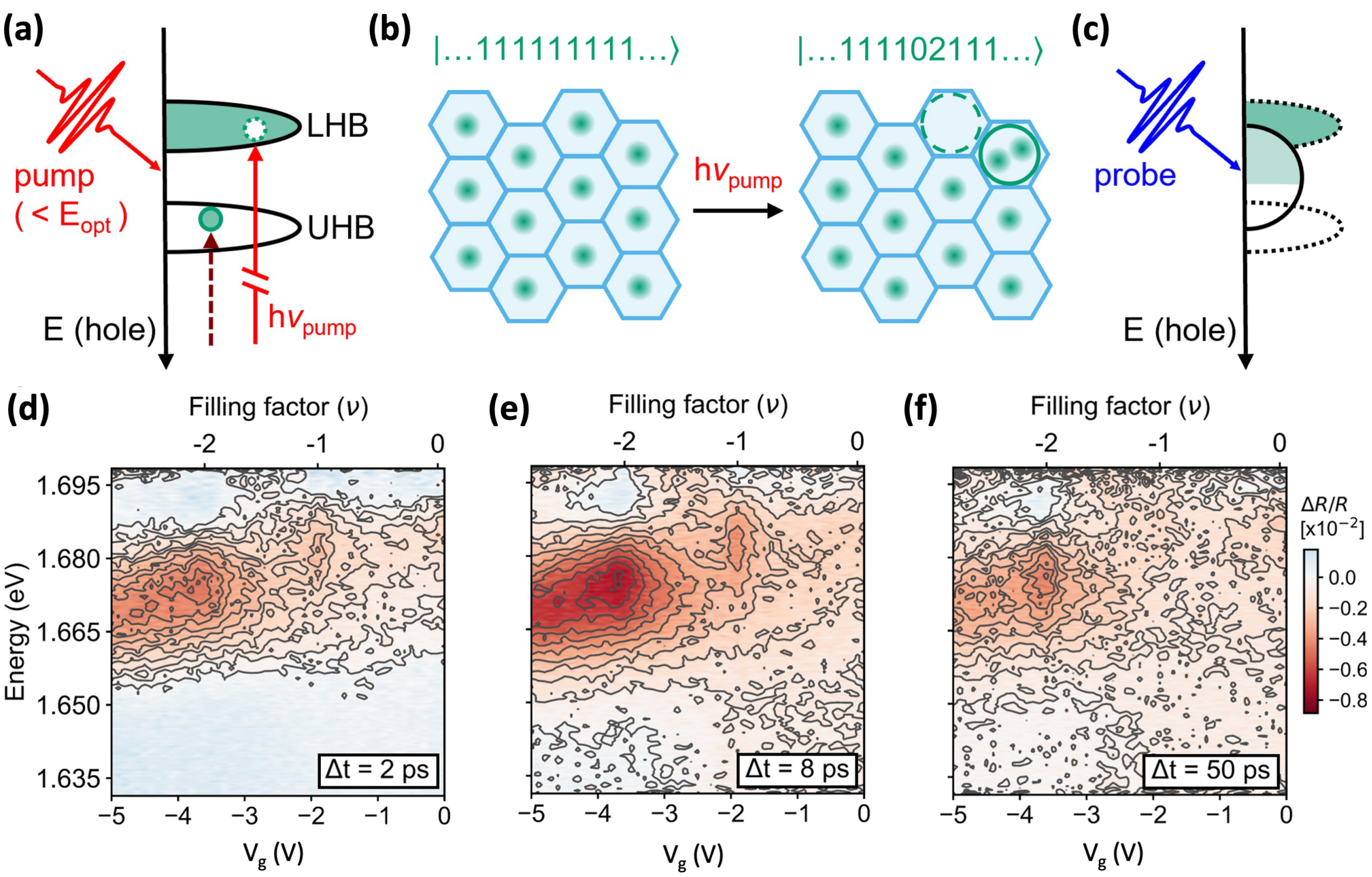

**Fig. 2. Pump-probe of a hole Mott state in hole-doped $WSe_2/WS_2$. a.** Excitation of a hole (green) from the lower Hubbard band (LHB) to a deep valence state, with subsequent cooling of the photo-excited hole (green sphere) to the upper Hubbard band (UHB). The hole energy scale is opposite to that of an electron scale. **b**. Real space representation of the holon-doublon pair for the hole (green spheres) Mott state, with corresponding state representation on the top of each cartoon; **c**. Partial melting (disordering) and gap closing from holon-doublon excitation. **d-f** Transient reflectance spectra plotted as a function of probe energy (covering the first moiré exciton of $WSe_2$) versus gate voltage ($V_g = V_t = V_b$) at Δt = 2, 8, 50 ps in d-f, respectively. Features specific to the $\nu$ = -1, -2 states can be observed and seen to evolve as a function of Δt. Here, *ΔR/R* is the transient reflectance signal calculated from pump-on and -off spectra. The pump fluence employed for the transient measurements was 57 $\mu J/cm^2$. All spectra collected at 11 K.

### Melting Dynamics of the Correlated States

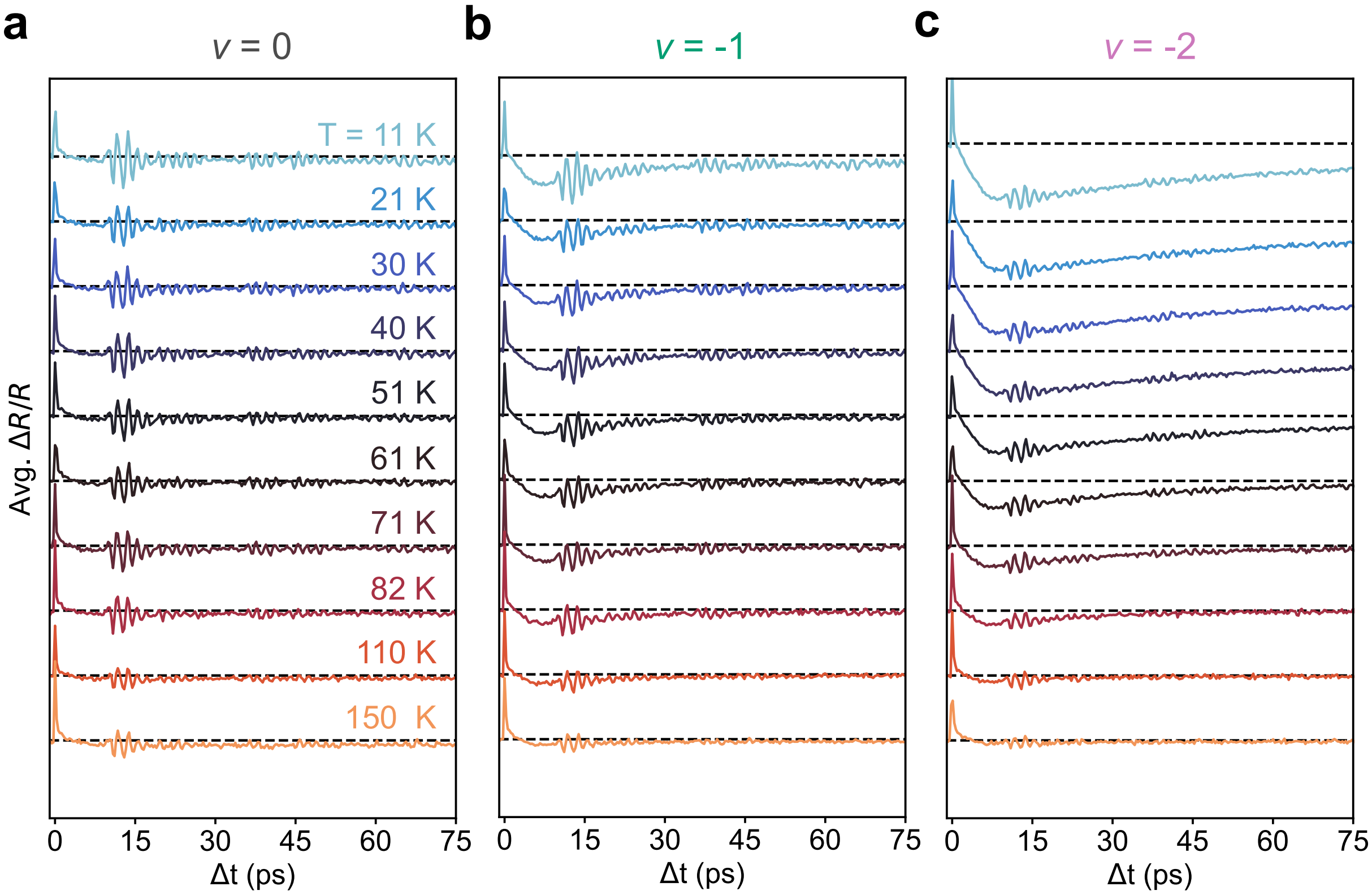

**Figure 3. Temperature-dependent temporal response of correlated states. a-c** Transient reflectance traces of the $v$ = 0, -1, -2 states, respectively, averaged over a ~0.01 eV probe energy range about the maximum of the lowest energy moiré band of $WSe_2$. The coherent artifact resulting from pump-probe overlap can be seen at Δt = 0 ps. The time traces have been offset for clarity. The colors corresponding to each temperature are labeled in panel a and are consistent throughout the figure. The employed pump fluence was 14 μJ/cm$^2$. All data shown is for sample D1.

Fig. 3 contains pump-probe data for the $v$ = 0, -1, and -2 states of the 60° sample. These measurements are carried out in a pump fluence regime where two photon absorption does not contribute (see Fig. S3 and Fig. S4). The only observed dynamics for $v$ = 0 (Fig. 3a) correspond to low frequency phonons, which appear as delayed and periodic modulations in $\Delta R/R$ on an otherwise flat background (see Figs. S5, S6). For the $v$ = -1, -2 states, in addition to phonons, we observe a decrease in overall $\Delta R/R$ in picoseconds, followed by recovery at longer times. The decrease in $\Delta R/R$ is attributed to an increase in $\varepsilon_{eff}$ from disordering, referred to as "melting". We melting time ($\tau_m$) from single exponential fits to the initial decays in $\Delta R/R$ in Fig. 3b, 3c (see Fig. S5 for details): $\tau_m$ = 3.2±0.2 ps ($v$ = -1) and 3.4±0.1 ps ($v$ = -2) at T ≤ 40 K. These $\tau_m$ values are at least five times longer than those predicted from the charge hopping integrals, but are in the range estimated from acoustic phonons. The pump-induced changes to $\varepsilon_{eff}$ are not resolved for fractional filling factors, Fig. S7.

For either $\nu$ = -1 or -2, $\tau_m$ remains constant until ~ 40 K, above which it decays monotonically, Fig. 4a. There are two regions: 1) at T ≤ 40 K, temperature independence is expected from e-e interaction; 2) at T > 40 K, e-ph interactions become important and melting is thermally activated. We plot melting rates, $k_m = 1/\tau_m$, with temperature, Fig. 4b-c for $\nu$ = -1, -2, respectively, and fit the data to the Arrhenius equation, $k_m \propto \exp[-E_a/k_B T]$, where $E_a$ is the activation energy and $k_B$ the Boltzmann constant. A constant offset is included to account for the temperature-independent contribution. These fits yield $E_a$ = 18±3 meV and 13±2 meV for the $\nu$ = -1 and -2, respectively.

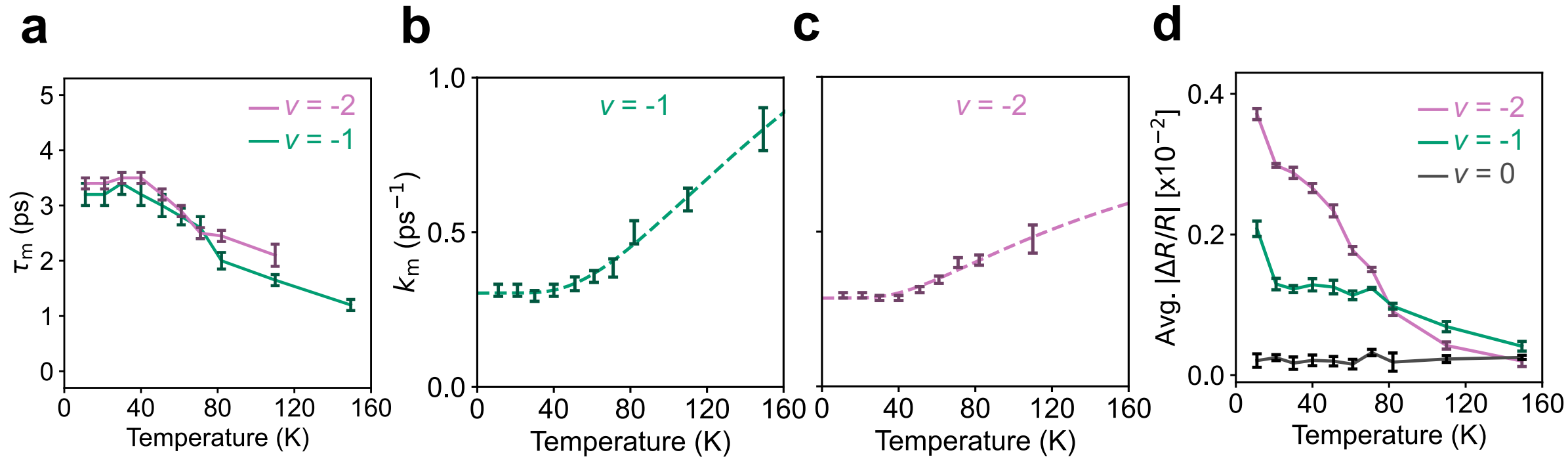

**Figure 4. Polaronic nature of correlated states.** **a** Temperature-dependent melting time ($\tau_m$) of the $\nu$ = -1, -2 states resulting from the pump pulse. **b, c** Melting rates, $k_m$, corresponding to panel **a**. The dashed lines are fits to the Arrhenius equation to extract thermal activation energies for $\nu$ = -1, -2 of $E_a$ = 18±3 meV and 13±2 meV, respectively. Constant offsets account for non-zero melting rates at lower temperatures. **d** Temperature-dependent transient reflectance for the $\nu$ = 0, -1, -2 states (shown in gray, green, purple, respectively). For each data point, the corresponding transient reflectance spectra was averaged over a ~0.01 eV probe energy range about the maximum of the lowest energy $WSe_2$ moiré exciton peak and averaged over a pump-probe delay of Δt = 4-8 ps. All error bars indicate 99% confidence intervals. All data shown is for sample D1.

The activation energies obtained from the melting rates are on the order of thermal dissociation energies estimated from $T_c$. While intensity of the sensing exciton in static reflectance (Fig. S8) may be used as a proxy to the order parameter in the Mott states [4,8], the presence of background signal makes analysis difficult. We use transient reflectance instead for $\nu$ = -1 (green) and -2 (purple), with $\nu$ = 0 (black) as control, Fig. 4d. The correlated states are still observed at ~150 K and ~100 K for $\nu$ = -1, -2, respectively, providing lower bounds for $T_c$. These give $k_b T_c$ ~ 13 meV and ~ 9 meV for $\nu$ = -1 and -2, respectively; they are ~70% of the $E_a$ values.

The temperature-dependent measurements are also repeated on a second 0° sample (D2) (Fig. S9). We obtain $\tau_m$ = 3.4±0.2 ps for $\nu$ = -1 at T = 8 K. The temperature dependence in $\tau_m$ at T > 40

K gives $E_a$ =19±3 meV and that of $\Delta R/R$ signal gives $T_c$ ~150 K for the $\nu$ = -1 state. These results are in excellent agreement with those measured in the 60° sample (D1) and confirm that the similar thermal dissociation temperatures observed for the correlated states in the 0° and 60° samples share the same physical origin. The differences in the moiré potential trap depth at 0° are 60° [24] do not seem to impact the melting/reordering dynamics of the Mott states. While we focus on melting, the recovery dynamics (Fig. S10) are more complex. The recovery of the $\nu$ = -1 state is essentially temperature-independent, suggesting that e-e scattering limits the reformation of the Mott state [25]. The recovery of the $\nu$ = -2 state, however, shows significant temperature dependence, pointing to the role of e-ph interactions. These processes deserve further investigations.

**Polaronic Nature of the Correlated Mott Insulators**

The melting rates and, more importantly, their thermal activations suggest electron-phonon coupling in the $\nu$ = -1 and -2 Mott states. In a correlated insulator, the residence time of each electron (hole) in a moiré unit cell is much longer than that of an uncorrelated electron, due to inhibition of charge hopping and much reduced kinetic energy in the former. The long residence time can lead to local polarization of the lattice to form a small polaron with the signature thermal activation [16]. The correlated insulator states are likely small polaron lattices. Here we use "small" to describe the size of the polaron with respect to the moiré length scale.

To corroborate polaron formation, we carry out density functional theory (DFT) calculations based on the PBE exchange-correlation functional on a $\theta$ = 0° $WSe_2/WS_2$ bilayer (Methods, Fig. S11). We first relax the structure of the neutral system, followed by additional structural relaxation upon introducing a positive charge into the moiré unit cell ($\nu$ = -1). Charge localization is evident from the spatial distribution of geometrical rearrangement which occurs upon injection of a positive charge into each unit cell, Fig. 5. The periodic electron density in the moiré superlattice introduces periodic lattice distortion, which is 90% perpendicular to the bilayer plane. We calculate a total reorganization energy of 16 meV; this can be taken as the polaron binding energy. This calculation is at the single-hole level, without the explicit inclusion of many-body correlation in the Hubbard Hamiltonian [3]. The resulting polaron binding energy agrees surprisingly well with experimental activation energies; although PBE-DFT is not expected to quantitatively account for e-ph coupling, it correctly describes the pattern of structural reorganization [26]. The disordering or melting of an ordered hole lattice requires hole hopping among moiré unit cells. In the small polaron picture, charge hopping requires thermal activation of phonons to overcome the polaron binding energy and release the hole from the trap. Thus, the thermal activation energy is related to the polaron binding energy [16,17,27].

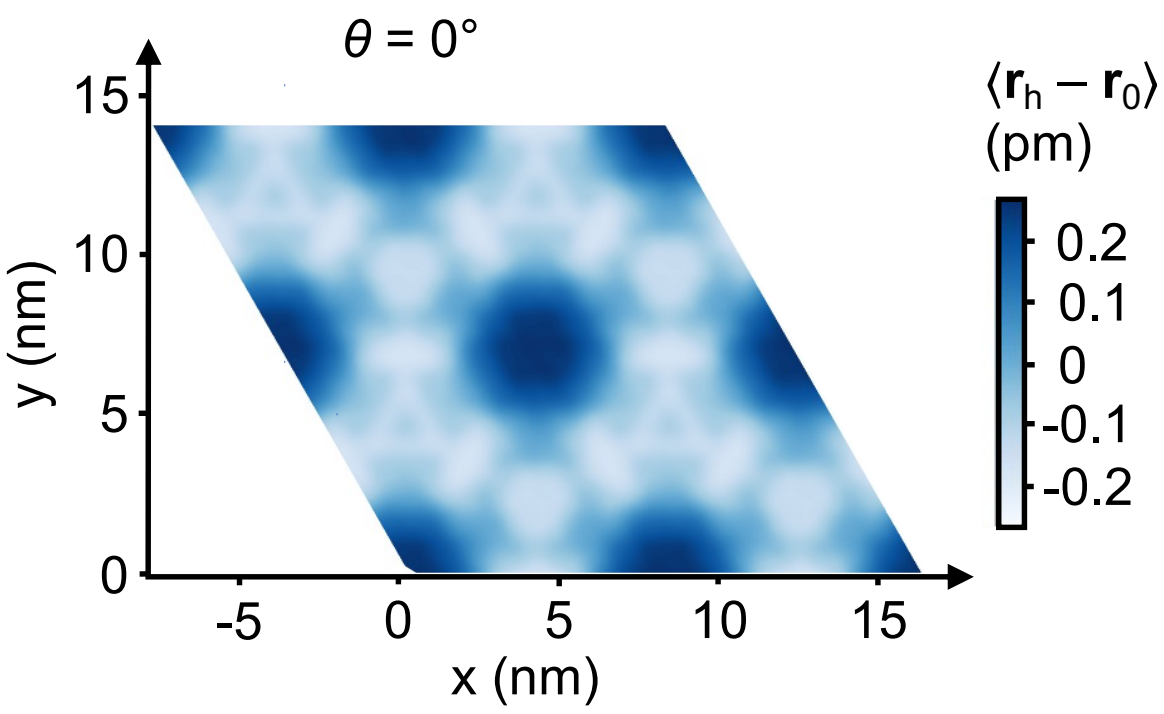

**Figure 5.** ***Ab initio*** **charge-induced lattice distortion.** Structural reorganization averaged over the atomic positions on each lattice point upon injection of a positive charge in an aligned $WS_2/WSe_2$ bilayer. The color scale represents the degree of distortion at each lattice position, $\langle \mathbf{r}_h - \mathbf{r}_0 \rangle$.

### Concluding Comments

The results presented here reveal the polaronic nature of correlated insulator states in the $WSe_2/WS_2$ moiré system. Our findings challenge the conventional many-body electron Hamiltonian and suggest the necessity of including e-ph coupling in some cases. In 2D materials, Kang et al. provided experimental evidence for Holstein "small" polaron in electron doped $MoS_2$ from intervalley coupling by zone-boundary longitudinal acoustic phonons [28]. Sio and Giustino reported polaron formation in h-BN; these authors suggested that Fröhlich "large" polaron formation from long-range Coulomb interaction is insignificant in 2D TMDs [29]. Recent theoretical work has also begun to explore e-ph coupling [30–32] in MATBG. The small polaron lattice reported here for Mott states in the TMD moiré system is similar to a CDW state and e-ph interactions may be partially responsible for thermal stability.

### References

[1] Y. Cao et al., *Correlated Insulator Behaviour at Half-Filling in Magic-Angle Graphene Superlattices*, Nature **556**, 80 (2018).

[2] Y. Cao, V. Fatemi, S. Fang, K. Watanabe, T. Taniguchi, E. Kaxiras, and P. Jarillo-Herrero, *Unconventional Superconductivity in Magic-Angle Graphene Superlattices*, Nature **556**, 43 (2018).

[3] F. Wu, T. Lovorn, E. Tutuc, and A. H. Macdonald, *Hubbard Model Physics in Transition Metal Dichalcogenide Moiré Bands*, Phys Rev Lett **121**, 26402 (2018).

[4] Y. Tang et al., *Simulation of Hubbard Model Physics in WSe2/WS2 Moiré Superlattices*, Nature **579**, 353 (2020).

[5] E. C. Regan et al., *Mott and Generalized Wigner Crystal States in $WSe_2/WS_2$ Moiré Superlattices*, Nature **579**, 359 (2020).

[6] L. Wang et al., *Correlated Electronic Phases in Twisted Bilayer Transition Metal Dichalcogenides*, Nat Mater **19**, 861 (2020).

[7] Y. Shimazaki, I. Schwartz, K. Watanabe, T. Taniguchi, M. Kroner, and A. Imamoğlu, *Strongly Correlated Electrons and Hybrid Excitons in a Moiré Heterostructure*, Nature **580**, 472 (2020).

[8] Y. Xu, S. Liu, D. A. Rhodes, K. Watanabe, T. Taniguchi, J. Hone, V. Elser, K. F. Mak, and J. Shan, *Correlated Insulating States at Fractional Fillings of Moiré Superlattices*, Nature **587**, 214 (2020).

[9] H. Li, S. Li, E. C. Regan, D. Wang, W. Zhao, S. Kahn, K. Yumigeta, M. Blei, T. Taniguchi, and K. Watanabe, *Imaging Two-Dimensional Generalized Wigner Crystals*, Nature **597**, 650 (2021).

[10] E. Liu, T. Taniguchi, K. Watanabe, N. M. Gabor, Y.-T. Cui, and C. H. Lui, *Excitonic and Valley-Polarization Signatures of Fractional Correlated Electronic Phases in a WSe2/WS2 Moiré Superlattice*, Phys Rev Lett **127**, 37402 (2021).

[11] X. Wang, C. Xiao, H. Park, J. Zhu, C. Wang, T. Taniguchi, K. Watanabe, J. Yan, D. Xiao, and D. R. Gamelin, *Light-Induced Ferromagnetism in Moiré Superlattices*, Nature **604**, 468 (2022).

[12] X. Huang, T. Wang, S. Miao, C. Wang, Z. Li, Z. Lian, T. Taniguchi, K. Watanabe, S. Okamoto, and D. Xiao, *Correlated Insulating States at Fractional Fillings of the WS2/WSe2 Moiré Lattice*, Nat Phys **17**, 715 (2021).

[13] K. Rossnagel, *On the Origin of Charge-Density Waves in Select Layered Transition-Metal Dichalcogenides*, Journal of Physics: Condensed Matter **23**, 213001 (2011).

[14] S. Hellmann et al., *Time-Domain Classification of Charge-Density-Wave Insulators*, Nat Commun **3**, 1069 (2012).

[15] K. F. Mak and J. Shan, *Semiconductor Moiré Materials*, Nat Nanotechnol **17**, 1 (2022).

[16] D. Emin, *Polarons* (Cambridge University Press, Cambridge, 2013).

[17] C. Franchini, M. Reticcioli, M. Setvin, and U. Diebold, *Polarons in Materials*, Nat Rev Mater **6**, 1 (2021).

[18] N. Mounet, M. Gibertini, P. Schwaller, D. Campi, A. Merkys, A. Marrazzo, T. Sohier, I. E. Castelli, A. Cepellotti, and G. Pizzi, *Two-Dimensional Materials from High-Throughput Computational Exfoliation of Experimentally Known Compounds*, Nat Nanotechnol **13**, 246 (2018).

[19] See Supplemental Material [Url] for additional results and analysis, which includes Refs. [33-41].

[20] F. Wu, T. Lovorn, E. Tutuc, and A. H. MacDonald, *Hubbard Model Physics in Transition Metal Dichalcogenide Moiré Bands*, Phys Rev Lett **121**, 026402 (2018).

[21] Y. Zhang, N. F. Q. Yuan, and L. Fu, *Moiré Quantum Chemistry: Charge Transfer in Transition Metal Dichalcogenide Superlattices*, Phys Rev B **102**, 201115 (2020).

[22] H. Park et al., *Dipole Ladders with Large Hubbard Interaction in a Moiré Exciton Lattice*, Nat Phys **19**, 1286 (2023).

[23] P. Rivera, H. Yu, K. L. Seyler, N. P. Wilson, W. Yao, and X. Xu, *Interlayer Valley Excitons in Heterobilayers of Transition Metal Dichalcogenides*, Nat Nanotechnol **13**, 1004 (2018).

[24] L. Yuan, B. Zheng, J. Kunstmann, T. Brumme, A. B. Kuc, C. Ma, S. Deng, D. Blach, A. Pan, and L. Huang, *Twist-Angle-Dependent Interlayer Exciton Diffusion in WS2–WSe2 Heterobilayers*, Nat Mater **19**, 617 (2020).

[25] N. Strohmaier, D. Greif, R. Jördens, L. Tarruell, H. Moritz, T. Esslinger, R. Sensarma, D. Pekker, E. Altman, and E. Demler, *Observation of Elastic Doublon Decay in the Fermi-Hubbard Model*, Phys Rev Lett **104**, 080401 (2010).

[26] K. Miyata, D. Meggiolaro, M. T. Trinh, P. P. Joshi, E. Mosconi, S. C. Jones, F. De Angelis, and X. -Y. Zhu, *Large Polarons in Lead Halide Perovskites*, Sci Adv **3**, e1701217 (2017).

[27] T. Holstein, *Studies of Polaron Motion: Part II. The "Small" Polaron*, Ann Phys (N Y) **8**, 343 (1959).

[28] M. G. Kang, S. W. Jung, W. J. Shin, Y. Sohn, S. H. Ryu, T. K. Kim, M. Hoesch, and K. S. Kim, *Holstein Polaron in a Valley-Degenerate Two-Dimensional Semiconductor*, Nat Mater **17**, 676 (2018).

[29] W. H. Sio and F. Giustino, *Polarons in Two-Dimensional Atomic Crystals*, Nat Phys 1 (2023).

[30] B. Lian, Z. Wang, and B. A. Bernevig, *Twisted Bilayer Graphene: A Phonon-Driven Superconductor*, Phys Rev Lett **122**, 257002 (2019).

[31] Y. W. Choi and H. J. Choi, *Dichotomy of Electron-Phonon Coupling in Graphene Moiré Flat Bands*, Phys Rev Lett **127**, 167001 (2021).

[32] M. Angeli, E. Tosatti, and M. Fabrizio, *Valley Jahn-Teller Effect in Twisted Bilayer Graphene*, Phys Rev X **9**, 041010 (2019).

[33] L. J. McGilly et al., *Visualization of Moiré Superlattices*, Nat Nanotechnol **15**, 580 (2020).

[34] M. R. Rosenberger, H.-J. Chuang, K. M. McCreary, A. T. Hanbicki, S. v Sivaram, and B. T. Jonker, *Nano-"Squeegee" for the Creation of Clean 2D Material Interfaces*, ACS Appl Mater Interfaces **10**, 10379 (2018).

[35] J. Wang, Q. Shi, E.-M. Shih, L. Zhou, W. Wu, K. Barmak, J. Hone, C. Dean, and X. Zhu, *Diffusivity Reveals Three Distinct Phases of Interlayer Excitons in MoSe2/WSe2 Heterobilayers*, Phys Rev Lett **126**, 106804 (2021).

[36] J. Wang, J. Ardelean, Y. Bai, A. Steinhoff, M. Florian, F. Jahnke, X. Xu, M. Kira, J. Hone, and X.-Y. Zhu, *Optical Generation of High Carrier Densities in 2D Semiconductor Heterobilayers*, Sci Adv **5**, eaax0145 (2019).

[37] Y. J. Bae et al., *Exciton-Coupled Coherent Magnons in a 2D Semiconductor*, Nature **608**, 282 (2022).

[38] J. Hutter, M. Iannuzzi, F. Schiffmann, and J. VandeVondele, *Cp2k: Atomistic Simulations of Condensed Matter Systems*, Wiley Interdiscip Rev Comput Mol Sci **4**, 15 (2014).

[39] J. VandeVondele, M. Krack, F. Mohamed, M. Parrinello, T. Chassaing, and J. Hutter, *Quickstep: Fast and Accurate Density Functional Calculations Using a Mixed Gaussian and Plane Waves Approach*, Comput Phys Commun **167**, 103 (2005).

[40] J. P. Perdew, K. Burke, and M. Ernzerhof, *Generalized Gradient Approximation Made Simple*, Phys Rev Lett **77**, 3865 (1996).

[41] D. G. A. Smith, L. A. Burns, K. Patkowski, and C. D. Sherrill, *Revised Damping Parameters for the D3 Dispersion Correction to Density Functional Theory*, J Phys Chem Lett **7**, 2197 (2016).

**Acknowledgements.** XYZ acknowledges support by DOE-BES under award DE-SC0024343. CRD acknowledges support for sample fabrication by the Materials Science and Engineering Research Center (MRSEC) through NSF grant DMR-2011738. XDX acknowledges support for sample fabrication and characterization by the Materials Science and Engineering Research Center (MRSEC) through NSF grant DMR-2011738. K.W. and T.T. acknowledge support from the JSPS KAKENHI (Grant Numbers 19H05790, 20H00354 and 21H05233). EAA gratefully acknowledges support from the Simons Foundation as a Junior Fellow in the Simons Society of Fellows (965526). ER acknowledges funding from the European Research Council (ERC) under the Horizon Europe Research and Innovation Program (Grant No. ERC-StG-2021-101040197—QED-Spin). We thank Di Xiao, Ting Cao, Liang Fu, Shiwei Zhang, Jordan Pack, Yinjie Guo for valuable help and discussions.